\documentclass [12pt]{article}
\usepackage {graphicx}
\usepackage[american]{babel}

\begin{document}

\title{Cooperative regulation of cellular identity in systems with intercellular communication defects}

\author{Nataliya~Stankevich$^{1,2}$ and Aneta~Koseska$^3$}

\maketitle
\begin{center}
\textit{$^1$Department of Applied Cybernetics, St. Petersburg State University, Saint-Petersburg, Russia}\\
\textit{$^2$Department of Radio-Electronics and Telecommunications, Yuri Gagarin State Technical University of Saratov, Saratov, Russia}\\
\textit{$^3$Department of Systemic Cell Biology, Max Planck Institute of Molecular Physiology, Dortmund, Germany}

\end{center}

\begin{abstract}
{The cooperative dynamics of cellular populations emerging from the underlying interactions determines cellular functions, and thereby their identity in tissues. Global deviations from this dynamics on the other hand reflects pathological conditions. However, how these conditions are stabilized from dis-regulation on the level of the single entities is still unclear. Here we tackle this question using generic Hodgkin-Huxley type of models that describe physiological bursting dynamics of pancreatic beta-cells, and introduce channel dis-function to mimic pathological silent dynamics. The probability for pathological behavior in beta-cell populations is $\sim100\%$ when all cell have these defects, despite the negligible size of the silent state basin of attraction for single cells. In stark contrast, in a more realistic scenario for a heterogeneous population, stabilization of the pathological state depends on the size of the sub-population which acquired the defects. However, the probability to exhibit stable pathological dynamics in this case is less than $\sim10\%$. These results therefore suggest that the physiological bursting dynamics of a population of beta-cells is cooperatively regulated, even under intercellular communication defects induced by dis-functional channels of single-cells.\\

Keywords: Beta-cells dynamics; disfunctional channels; heterogeneous cell population.\\
\\ }
\end{abstract}


\section{INTRODUCTION}

The emergent properties of complex biological systems are largely determined by the interactions of the elements that constitute these systems. Synchronous firing of ensembles of interacting neurons enables for example, complex collective tasks such as Hebbian learning and memory \cite{a1}, \cite{a2} or wake-sleep cycles \cite{a3}, \cite{a4} to be executed. On the other hand, the collective behavior of homogeneous or heterogeneous cell types including their arrangement and types of contacts determines specific functions of tissues and organs. For instance, it has been experimentally observed that the well-synchronised electrical activity (spiking and bursting) of primary beta-cells in the pancreatic islet is what precedes insulin secretion \cite{a5}, \cite{a6}, \cite{a7}, where the secretion increases with the fraction of time that the cells spend in the spiking state. The duration of the silent phase between two bursts is in this case regulated by the rate at which calcium is removed from the cell interior. The spiking oscillations typically display a period of 1-10s, whereas the duration of the bursting period varies from 0.2 to 5min. In contrast, isolated cells show electrical activity with very different time scales, do not burst and thereby do not secrete insulin effectively \cite{a8}, \cite{a9}. In this case, the opening probability of the potassium channels is too small for the individual cells to present  regular spiking signal that conversely only occurs in population of synchronized cells, thereby demonstrating that the intercellular communication in the organized islet structure is a necessary prerequisite for effective functioning of this coupled-cell system. Deviation from the characteristic dynamical behavior of the system on the other hand reflects pathological conditions. How such global deviations in the dynamics are stabilized from dis-regulation on the level of the interacting elements is however unclear. This question becomes even more prominent considering that it has been experimentally demonstrated that generically, the collective behavior of cells on the level of tissues is generally stable against mutations that occur in single cells, allowing them to retain their identity \cite{a10}.

Here we use numerical simulations to tackle how a change in the global dynamics of a coupled system can be established by dis-regulation on a sub-population level. To address this, we employed a generic Hodgkin-Huxley formalism which on a single cell level describes physiological bursting activity of pancreatic beta-cells \cite{a11}, \cite{a12}, \cite{a13}. Additionally, we also use a modified version of this model where a potassium-like ion channel with decreased opening probability is introduced \cite{a14} to stabilize a silent state in addition to the bursting activity, thereby enabling bistability on a single-cell level to be established. This silent state mimics ion channel disfunction in a form of blocking or inactivation. We use these models to investigate how deviation from the physiological dynamics of the population of coupled beta-cells is triggered depending on the number of cells acquiring disregulation in channel activity. Three cases are therefore analyzed: (1) dynamics of two coupled beta-cells under physiological conditions (bursting activity), (2) dynamical behavior of two coupled cells that have acquired the dis-function (stable steady state), but one or both of them do not manifest it in isolated cells and (3) dependence of the global population dynamics on number of cells with channel defects.
We found that the dynamics of a population of two coupled cells, each characterized with bistability between silent and bursting state, is either equivalent to the same attractor in which both cells are poised when in isolation, or dictated by the physiological bursting dynamics when the uncoupled cells are initially poised in different attractors. In contrast, the bursting dynamics of a heterogeneous population in which only a portion of cells can exhibit bistability between silent and bursting dynamics and are subsequently poised in the silent state, can be only disrupted if at least $N/2+1$ cells have the channel dis-function. The probability for the full population to exhibit silent dynamics in this case is however less than $10\%$. These results thereby demonstrate that the identity of a heterogeneous population of beta-like cells reflected through their bursting dynamics is cooperatively regulated, even when a large sub-fraction of this population acquired dis-functional channels that induce intercellular communication defects.

\section{Results}

\subsection{Single beta-cell models of bursting and silent dynamics}

The electrical activity of pancreatic beta-cells relies on multiple types of voltage- and ligand-gated ion channels that are permeable to inorganic ions such as sodium, potassium, chloride and calcium \cite{a5},\cite{a15}, \cite{a16}. These channels not only regulate membrane potential, ion homeostasis and electrical signaling \cite{a17}, but a dis-regulation in their activity has been recently also implicated in tumorogenesis and tumor progression \cite{a18}, \cite{a19}, \cite{a20}, \cite{a21}.

\begin{figure}
\centerline{
\includegraphics[scale=0.7]{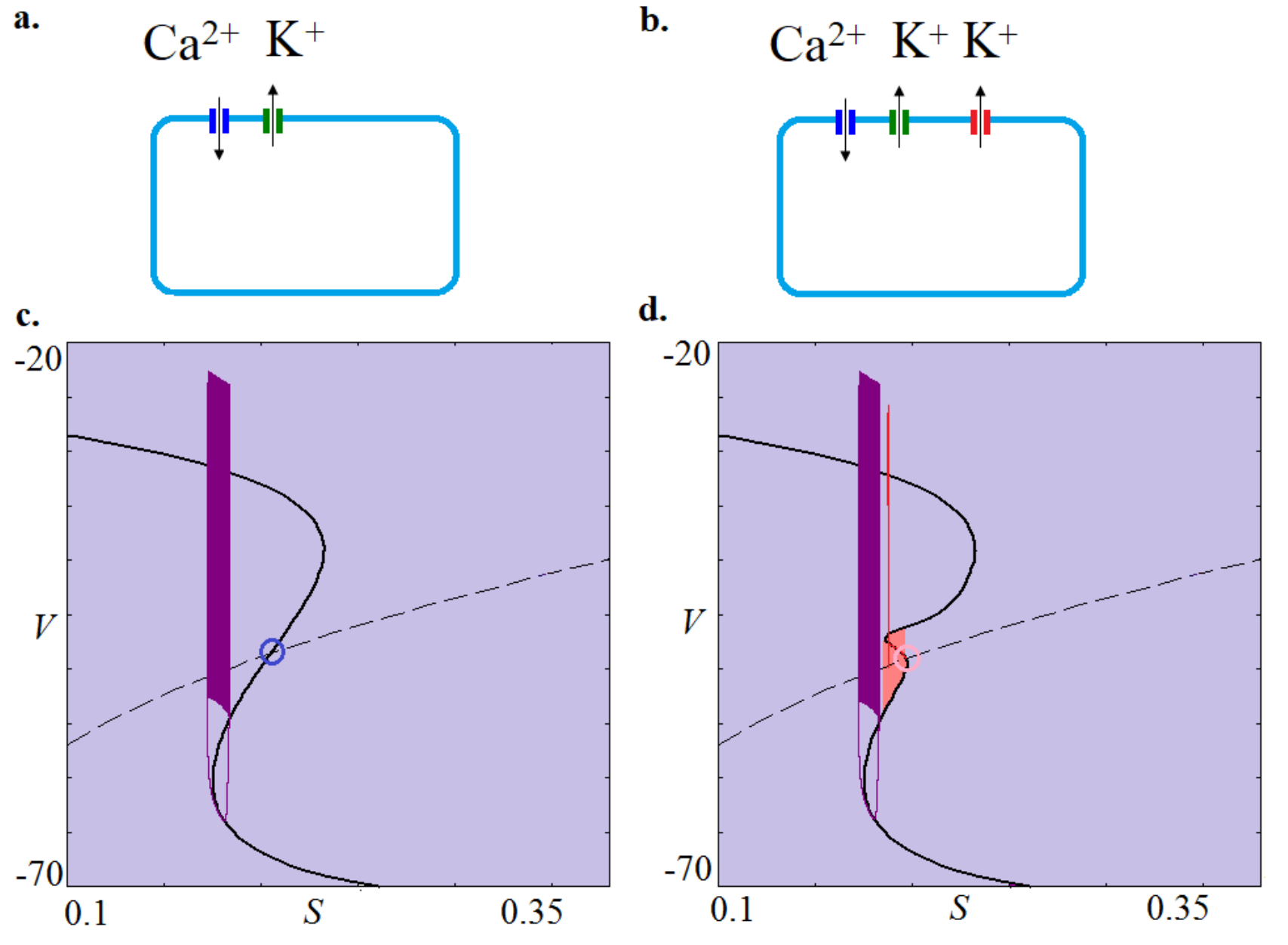}\\}
\caption{\textbf{Dynamical structure of the beta-cell models (Eqs. \ref{eq:main_model}-\ref{eq:probab1}).}
 Schematic representation of a beta-cell in {\bf a.} physiological conditions (blue: Ca-channel, green:K-channel with physiological opening probability) and {\bf b.} with an additional potassium channel with decreased opening probability (red). Fast and slow manifolds (black solid and dashed lines, respectively), attractors and their basins of attraction (blue: bursting attractor, red: stable node) are shown for {\bf c.} $k^{(i)}=0$, corresponding to Fig.~\ref{fig:fig1}a and {\bf d.} $k^{(i)}=1$, Fig.~\ref{fig:fig1}b. Stable and unstable nodes are marked with red and blue circles, respectively. Parameters:
$\tau = 0.02$, $\tau_{S} = 35$, $\sigma = 0.93$, $g_{Ca} = 3.6$, $g_{K} = 10.0$, $g_{K2} = 0.2$, $g_{S} = 4.0$, $V_{Ca} = 25.0$, $V_{K} = -75.0$, $V_m = -20.0$, $\theta_m = 12.0$, $V_n = -16.0$, $\theta_n = 5.6$, $V_S = -35$, $\theta_S = 10.0$, $V_P = -47.0$, $\theta_p = 1.0$.}
\label{fig:fig1}
\end{figure}

To depict the bursting dynamics of pancreatic beta-cells as observed under physiological conditions, as well as to simultaneously model silent dynamics caused upon ion channel dis-regulation, we use a model based on the Hodgkin-Huxley formalism \cite{a12}, \cite{a13} with modifications proposed by Stankevich {\it et al.} \cite{a14}:

\begin{equation}
\begin{array}{l l l}
\tau \frac{dV^{(i)}}{dt} = -I_{C_a}(V^{(i)}) - I_K(V^{(i)},n^{(i)}) - k^{(i)}I_{K2}(V^{(i)})\\[8pt]
-I_S(V^{(i)},S^{(i)}) \\[8pt]\\
\tau \frac{dn^{(i)}}{dt} = \sigma (n_{\infty}(V^{(i)}) - n^{(i)}) \\[8pt]
\tau_S \frac{dS^{(i)}}{dt} = S_{\infty}(V^{(i)}) - S^{(i)}
\end{array}
\label{eq:main_model}
\end{equation}

where $V^{(i)}$ represents the membrane potential of the {\it i-th} cell, the functions $I_{C_a}(V^{(i)})$, $I_K(V^{(i)},n^{(i)})$, $I_S(V^{(i)}, S^{(i)})$ define the three intrinsic currents, fast calcium, potassium and slow potassium respectively, such that:

\begin{equation}
I_{C_a}(V^{(i)}) = g_{C_a}m_{\infty}(V^{(i)})(V^{(i)}-V_{C_a})
\end{equation}
\begin{equation}
I_{K}(V^{(i)},n^{(i)}) = g_{K}n(V^{(i)}-V_{K})
\label{eq:normal_current}
\end{equation}
\begin{equation}
I_{S}(V^{(i)},n^{(i)}) = g_{S}S(V^{(i)}-V_{K})
\end{equation}

The gating variables for $n^{(i)}$ and $S^{(i)}$ are the opening probabilities of the fast and slow potassium currents:

\begin{equation}
\omega_{\infty}(V^{(i)}) = [1+exp\frac{V_{\omega}-V^{(i)}}{\Theta_{\omega}}]^{-1}, \omega=m,n,S
\label{eq:probab}
\end{equation}

The function $I_{K2}(V^{(i)})$ on the other hand defines an additional voltage-dependent potassium current. It has been previously demonstrated that a single oscillator, in absence of $I_{K2}$ (when $k^{(i)}=0$, Fig.~\ref{fig:fig1}a), is characterized with a bursting attractor that is born via a Hopf bifurcation for $V_S=-44.7mV$. At this parameter value, the equilibrium point looses its stability \cite{a22}, \cite{a23}. Although the bursting attractor is born in the vicinity of the equilibrium point, this point moves away from the bursting attractor for increase in $V_S$. Even more, the two dimensional projection of the phase portrait together with the fast and slow manifolds when $V_S=-35mV$ (Fig.~\ref{fig:fig1}c) demonstrates that the periodic trajectories do not intersect the neighborhood of the steady state and the bursting attractor terminates in a homoclininc bifurcation as the trajectory hits the slow manifold \cite{a22}, \cite{a23}. Calculating the eigenvalues in this case shows that the equilibrium point is an unstable node $(\lambda_1, \lambda_2, \lambda_3) = (23.14, -41.83, 0.09)$.

In the presence of $I_{K2}(V^{(i)})$ ($k^{(i)} \neq 0$, Fig.~\ref{fig:fig1}b) that varies strongly with the membrane potential in the vicinity of the equilibrium point however, the unstable node is stabilized without affecting the global flow of the model (Fig.~\ref{fig:fig1}d, corresponding eigenvalues $(\lambda_1, \lambda_2, \lambda_3) = (-37.58, -13.6, -0.24)$, estimated for $g_{K2}=0.2$ at the equilibrium point ($V_0^{(1)}$, $n_0^{(1)}$, $S_0^{(1)}$) = (-49.084, 0.0027105, 0.19648)) \cite{a14}. $I_{K2}(V^{(i)})$ is given in the form:

\begin{equation}
I_{K2}(V^{(i)}) = g_{K2}p_{\infty}(V^{(i)})(V^{(i)}-V_K)
\label{eq:new}
\end{equation}

where

\begin{equation}
p_{\infty}(V^{(i)})=[exp\frac{V^{(i)}-V_p}{\Theta_p}+exp\frac{V^{(i)}_p-V}{\Theta_p}]^{-1}
\label{eq:probab1}
\end{equation}

represents the opening probability of this channel. In contrast to the other potassium channel (Eqs. \ref{eq:normal_current}, \ref{eq:probab}) which opens with probability $n_{\infty}=1$ when the membrane voltage reaches a threshold value, the opening probability of the modified channel will be equal only to 0.5. From physiological point of view, this can be interpreted as an ion channel disfunction, as for instance blocking of the potassium channel or its inactivation \cite{a24}, and thereby a stable silent state emerges. Between the stable node and the bursting attractor there is a rejecting current which enables the system to remain in the stable steady state when starting from initial conditions in its vicinity \cite{a14}. Generally, the modified model ($k^{(i)} \neq 0$) depicts bistability between physiological, bursting beta-cell dynamics and a pathological, silent state dynamics (Fig.~\ref{fig:fig1}d).

\subsection{Dynamics of two coupled beta-cells under physiological conditions}

\begin{figure}
\centerline{
\includegraphics[scale=0.7]{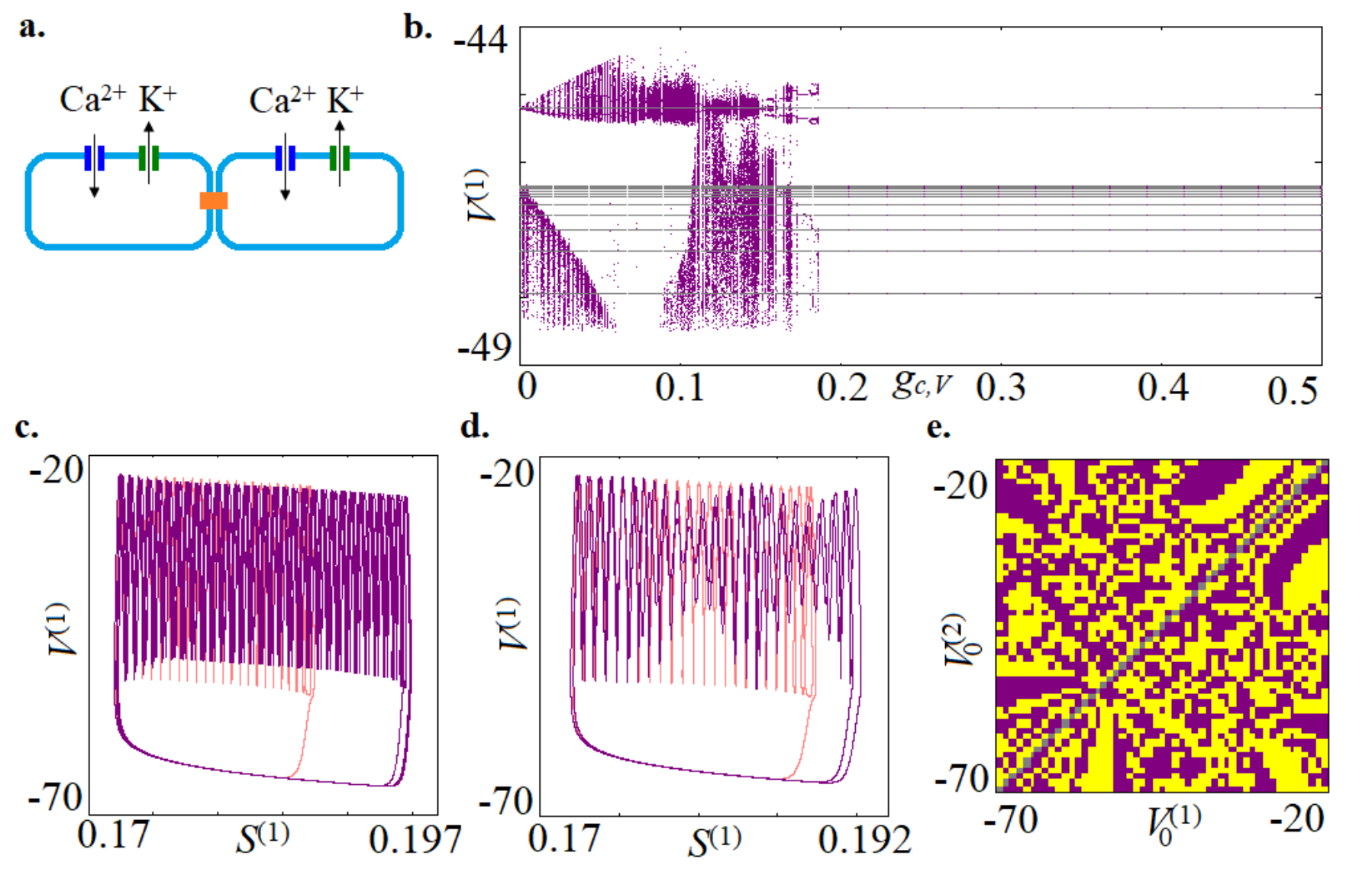}\\}
\caption{\textbf{Behavior of two coupled beta-cells with bursting dynamics.}
 {\bf a.} Schematic representation of the model. {\bf b.} Bifurcation trees estimated from Poincar\'{e} sections at the hypersurface $n^{(1)}=0.003$ when $g_{c,V}$ is scanned in both directions (violet: forward, gray: backward). $k^{(i)}=0, i=1,2$ and other parameters as in Fig.~\ref{fig:fig1}. Exemplary phase portraits depicting coexistence between synchronized bursting and {\bf c.} periodic or {\bf d.} chaotic attractors are also shown for $g_{c,V}=0.03$ and $0.15$ respectively. {\bf e.} Estimated basins of attraction of the coexisting bursting attractors for $g_{c,V}=0.15$ when starting from the following initial conditions: $n^{(1)}_{0} = 0.003$, $S^{(1)}_{0} = 0.1889747$, $n^{(2)}_{0} = 0.0025039$, $S^{(2)}_{0} = 0.1889565$.
}
\label{fig:fig2}
\end{figure}

To investigate the global dynamics of a population of beta-cells, we introduced electrical coupling by adding the following gap-junctional coupling term to the equation for $V^{(i)}$ that describes bi-directional transport of ions between the cells \cite{a7}, \cite{a25}:

\begin{equation}
I_C(V^{(i)})=\sum_{j \in \Gamma_i}g_{c,V}(V^{(i)}-V^{(j)})
\end{equation}

with $g_{c,V}$ being the coupling conductance (coupling strength). We consider only electrical coupling of the cells (Fig.~\ref{fig:fig2}a) by adding the following gap-junctional coupling terms to the equations for $V$ in (Eqs. \ref{eq:main_model}). The sum is taken over the whole population, assuming global coupling.

As already noted, if $k_i=0$ and in absence of coupling, the isolated oscillators display a single stable state - bursting dynamics (Fig.~\ref{fig:fig1}c). To infer the effect of coupling on the system's dynamics, we investigated next the bifurcation structure of a minimal model of two coupled cells as a function of the coupling strength $g_{c,V}$. Bifurcation trees using the membrane voltage variable $V^{(1)}$ were therefore constructed using a Poincar$\acute{e}$ section at the hypersurface $n^{(1)}=0.003$ by scanning $g_{c,V} \in [0, 0.6]$ from left to right and vice versa (Fig.~\ref{fig:fig2}b violet and grey, respectively).
Changing the scanning direction of the bifurcation parameter unraveled that for $g_{c,V} \in [0, 0.19]$, there is a coexistence between a synchronized bursting state and an additional dynamical regime. Numerical simulations for distinct coupling values from this interval showed that this dynamical regime could be periodic or chaotic (Fig.~\ref{fig:fig2}c,d respectively). This is also reflected in the complex fractal structure of the respective basins of attraction, as exemplified for $g_{c,V}=0.15$ in Fig.~\ref{fig:fig2}e. Given that the synchronized bursting state lies only on the diagonal of the basin (grey line in Fig.~\ref{fig:fig2}e), it is most probable to reach one of the other dynamical states in this parameter region. For $g_{c,V} > 0.19$ however, only synchronised bursting between both oscillators was observed.  This analysis therefore demonstrates that the physiological, synchronized bursting behavior is the dominant dynamical behavior over a wide coupling range in a population of identical beta-cells that communicate via gap junctions.

\subsection{Dynamics of two coupled beta-cells that contain dis-functional potassium channels}

\begin{figure}
\centerline{
\includegraphics[scale=0.5]{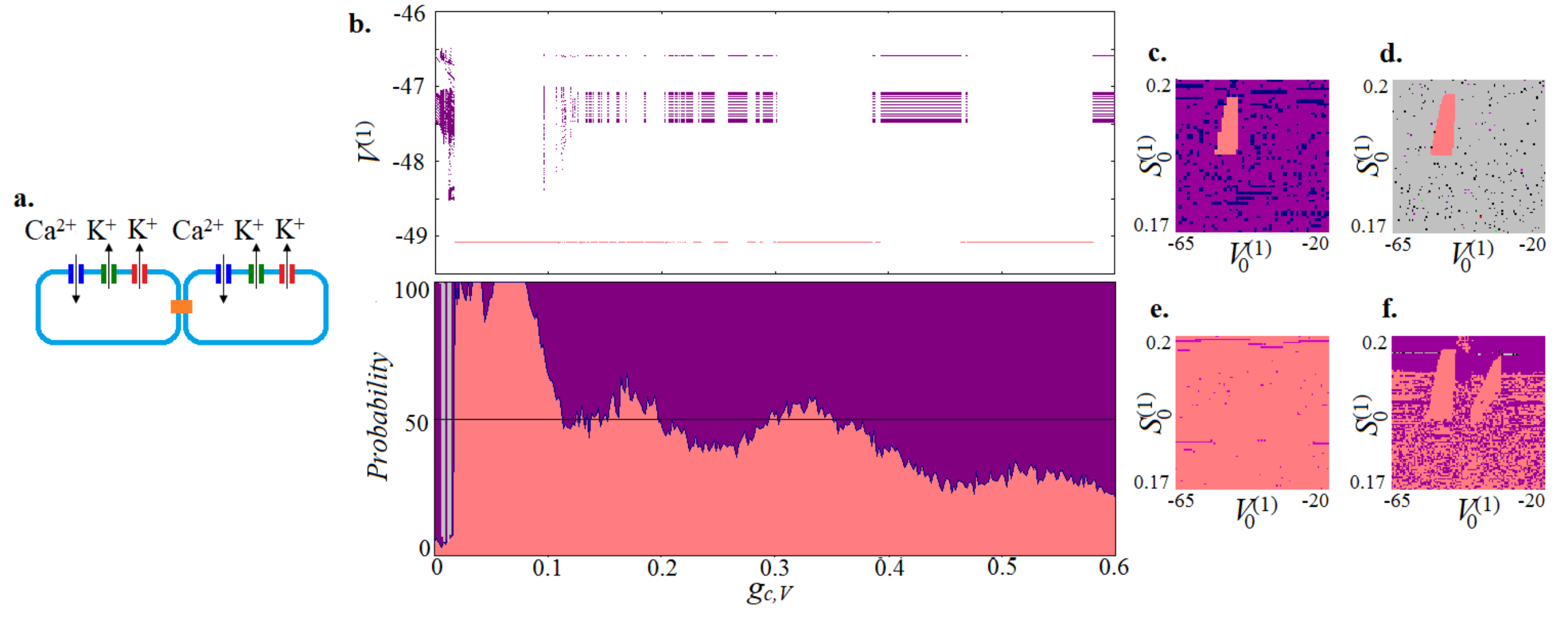}\\}
\caption{\textbf{Behavior of two coupled beta-cells with disfunctional potassium channels.}
 {\bf a.} Schematic representation of two the system. {\bf b.} (Top) Bifurcation tree estimated from Poincar\'{e} sections at the hypersurface and $n^{(1)}=0.003$ constructed for fixed initial conditions: $V^{(1)}_0 = -50.784$, $n^{(1)}_0 = 0.0245$, $S^{(1)}_0 = 0.1774$, $V^{(2)}_0 = -49.084$, $n^{(2)}_0 = 0.027105$, $S^{(2)}_0 = 0.19648$, as well as (bottom) probability of appearing of co-existing attractors in dependence on the coupling strength $g_{c,V}$. $k^{(i)}=1, i=1,2$ and other parameters as in Fig.~\ref{fig:fig1}. Exemplary basins of coexisting attractors for various strength of coupling: {\bf c.} $g_{c,V}=0.004$; {\bf d.} $g_{c,V}=0.00663$; {\bf e.} $g_{c,V}=0.02$; {\bf f.} $g_{c,V}=0.12$. For {\bf c.-f.}, the initial conditions of the other variables were fixed in steady state $n^{(1)}_0 = 0.027105$, $V^{(2)}_0 = -49.084$, $n^{(2)}_0 = 0.027105$, $S^{(2)}_0 = 0.19648$. Grey: chaotic dynamics; Violet: bursting dynamics; Red: steady state.
}
\label{fig:fig3}
\end{figure}

For $k_i=1$ however, the presence of the potassium channel which has a decreased opening probability (Fig.~\ref{fig:fig1}b, Eqs. \ref{eq:new},\ref{eq:probab1}) ensures that the equilibrium point is also stabilized, such that isolated oscillators display bistability between a bursting and a silent state in absence of coupling (Fig.~\ref{fig:fig1}d). Thus, for initial conditions in the vicinity of the stable node, the physiological bursting behavior is not observed. We therefore refer to this silent regime as pathological. To unravel how such channel defects affect the global dynamics of the population, we used again a minimal model of two coupled beta-cells that are individually characterized with bistability between the two states (Fig. ~\ref{fig:fig3}a), and investigated the two possible scenarios: {\it i)} the initial conditions of both oscillators are poised either in the bursting attractor or in the stable node attractor, and {\it ii)} one of the oscillators is poised in the stable node, whereas the other oscillator is poised in the bursting attractor.

When the initial conditions of both oscillators are poised in the same attractor, the solution of the coupled system is trivial - the population behavior is synchronized and identical to the dynamics of the cells in isolation (results not shown). In contrast, for scenario {\it ii)}, when isolated cells populate distinct attractors such that the initial conditions for one cell are poised in the bursting whereas for the other in the stable node, the bifurcation tree constructed analogously to the one in Fig.~\ref{fig:fig2}b has a complex structure for varying coupling strength $g_{c,V}$ (Fig.~\ref{fig:fig3}b, top). To identify the existing dynamical solutions in this case, we complemented the bifurcation diagram with direct calculations from semi-random initial conditions (Fig.~\ref{fig:fig3}b, bottom). In particular, for each $g_{c,V}$ we calculated 500 time series for the system of two coupled beta-cells such that the initial conditions for one of the oscillators were chosen exactly at the stable node $(V_0^{(1)}, n_0^{(1)}, S_0^{(1)}) = (-49.084, 0.0027105, 0.19648)$, whereas the initial conditions for the second oscillator were randomly chosen from a phase space cube in the interval $V_0^{(2)} \in (-60, -20)$, $n_0^{(2)} \in (0, 0.12)$ and $S_0^{(2)} \in (0.17, 0.2)$. Although the size of this cube  includes both, the bursting as well as the stable node attractor in the uncoupled case, the basin of the bursting one is dominant. The full set of initial conditions covers the 6-dimensional phase space of the system with 3 degrees of freedom per oscillator densely enough such that one can detect all stable coexisting attractors with a significant basin of attraction. These direct calculations represent a different approach from the bifurcation analysis, and are valid for large system sizes as well \cite{a26}. Therefore, the combination of both methods sheds light on the dynamics of the coupled beta-cells model under the introduced pathological channel defects.

The bifurcation tree and the direct calculations plot (Fig.~\ref{fig:fig3}b) demonstrate that under weak coupling ($g_{c,V}<0.017$), the system of two coupled beta-cells where initial conditions differentially poise the oscillators in the two stable attractors, predominantly displays either bursting or complex, chaotic dynamic. For the $g_{c,V}$ sub-intervals for which bursting or chaotic dynamics was observed, the probability for the coupled system to exhibit steady state solution was less than $10\%$. That the bursting or the chaotic dynamics are the dominant regimes under weak coupling is also reflected in the corresponding basins of attraction estimated for the respective $g_{c,V}$ values (Fig.~\ref{fig:fig3}c,d). In contrast, for a large part of intermediate coupling strength interval ($0.017<g_{c,V}<0.08$), the probability that the coupled system exhibits steady state dynamics was $\sim100\%$, with less than $10\%$ variability in small $g_{c,V}$ ranges. The exemplary basin of attraction plot estimated for $g_{c,V}=0.02$ also depicts the dominance of the stable equilibrium point, with only small lines of the basin of the bursting attractor (Fig.~\ref{fig:fig3}e). This is surprising, given the small basin of attraction of the steady state for isolated cells (Fig.~\ref{fig:fig1}d). The mechanism leading to this stabilization will be discussed elsewhere (manuscript in preparation).

A further increase in $g_{c,V}$ continuously decreases the probability for the full system to exhibit steady state behavior. At $g_{c,V} \approx 0.12$ for example, the system visits the bursting attractor and the equilibrium point with equal probability. This implies that in the presence of noise, the full system can switch between bursting physiological and silent pathological state.
For high enough coupling strength $g_{c,V}$ however, the basin of attraction of the steady state is significantly decreased, leaving the bursting attractor as the most common solution of the system (Fig.~\ref{fig:fig3}f).
These results therefore demonstrate that a system of two coupled beta-cells that have disfunctional potassium channels and thereby possibility to manifest a stable steady state, predominantly exhibits physiological bursting dynamics. This is true for weak and strong coupling strengths, even when the initial conditions of one of the oscillators correspond to its positioning in the stable node. Only for intermediate coupling values the silent pathological state is the dominant dynamical regime, even for small steady state basin of attraction for single cells.

\subsection{Dynamical behavior of a minimal mixed population model}

\begin{figure}
\centerline{
\includegraphics[scale=0.5]{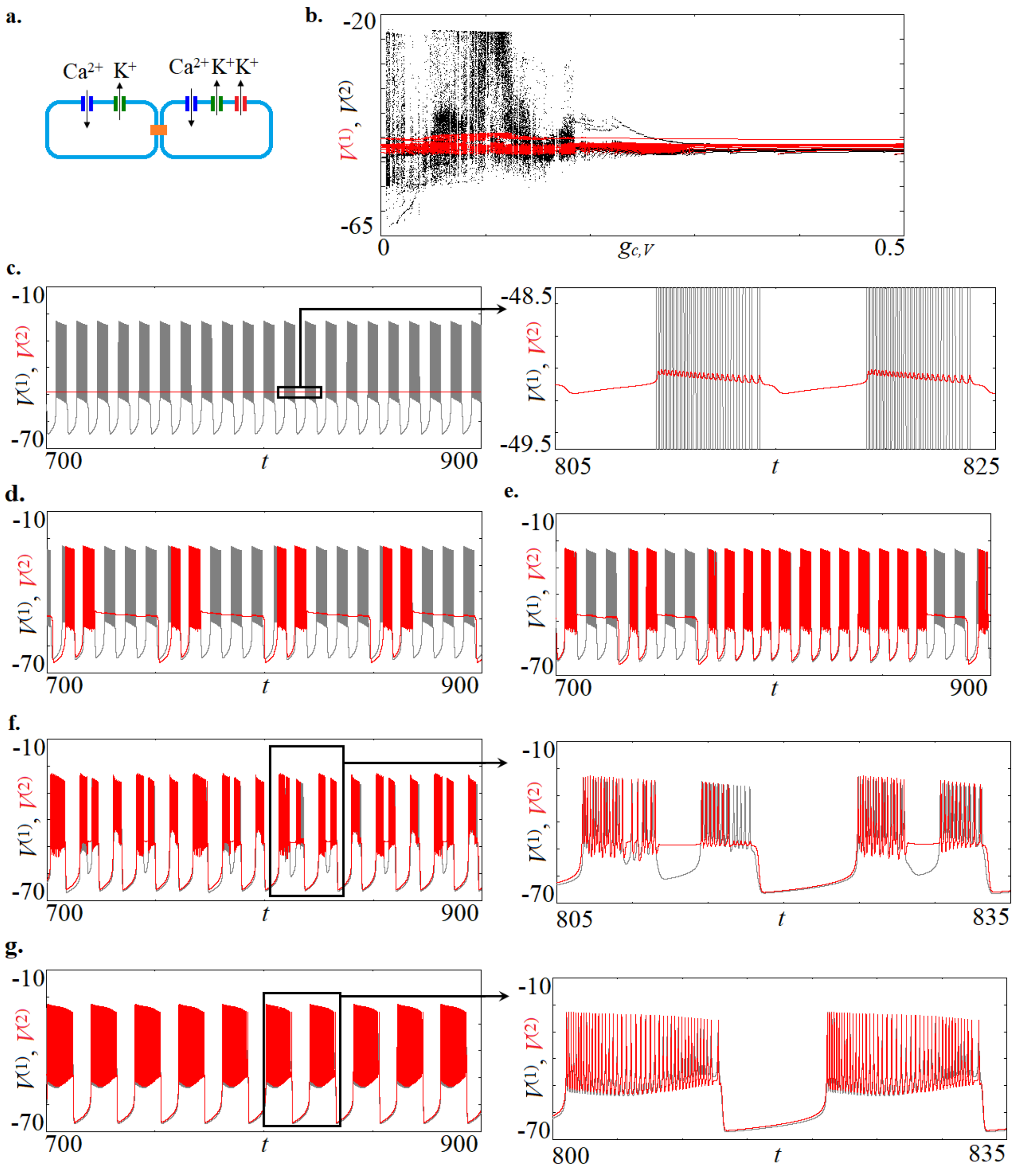}\\}
\caption{\textbf{Behavior of a minimal mixed population model.}
 {\bf a.} Schematic representation of minimal population of two coupled heterogeneous cells. {\bf b.} Bifurcation tree estimated from Poincar\'{e} sections at the hypersurface $n^{(1)}=0.003$ when $g_{c,V}$ is scanned in both directions. $k^{(1)}=0, k^{(2)}=1$ and other parameters as in Fig.~\ref{fig:fig1}.
 Exemplary time series of characteristic dynamical regimes for: {\bf c.} $g_{c,V}=0.001$ {\bf d.} $g_{c,V}=0.008$; {\bf e.} $g_{c,V}=0.015$; {\bf f.} $g_{c,V}=0.07$; {\bf g.} $g_{c,V}=0.3$.
}
\label{fig:fig4}
\end{figure}

To model physiologically more realistic scenario, we investigate next the dynamics of a minimal mixed system of two coupled cells, where only one of them displays bistability between bursting and a silent state, and thus a dis-regulation in a form of inactivation or blocking of a potassium channel (Fig.~\ref{fig:fig4}a).

\begin{figure}
\centerline{
\includegraphics[scale=0.5]{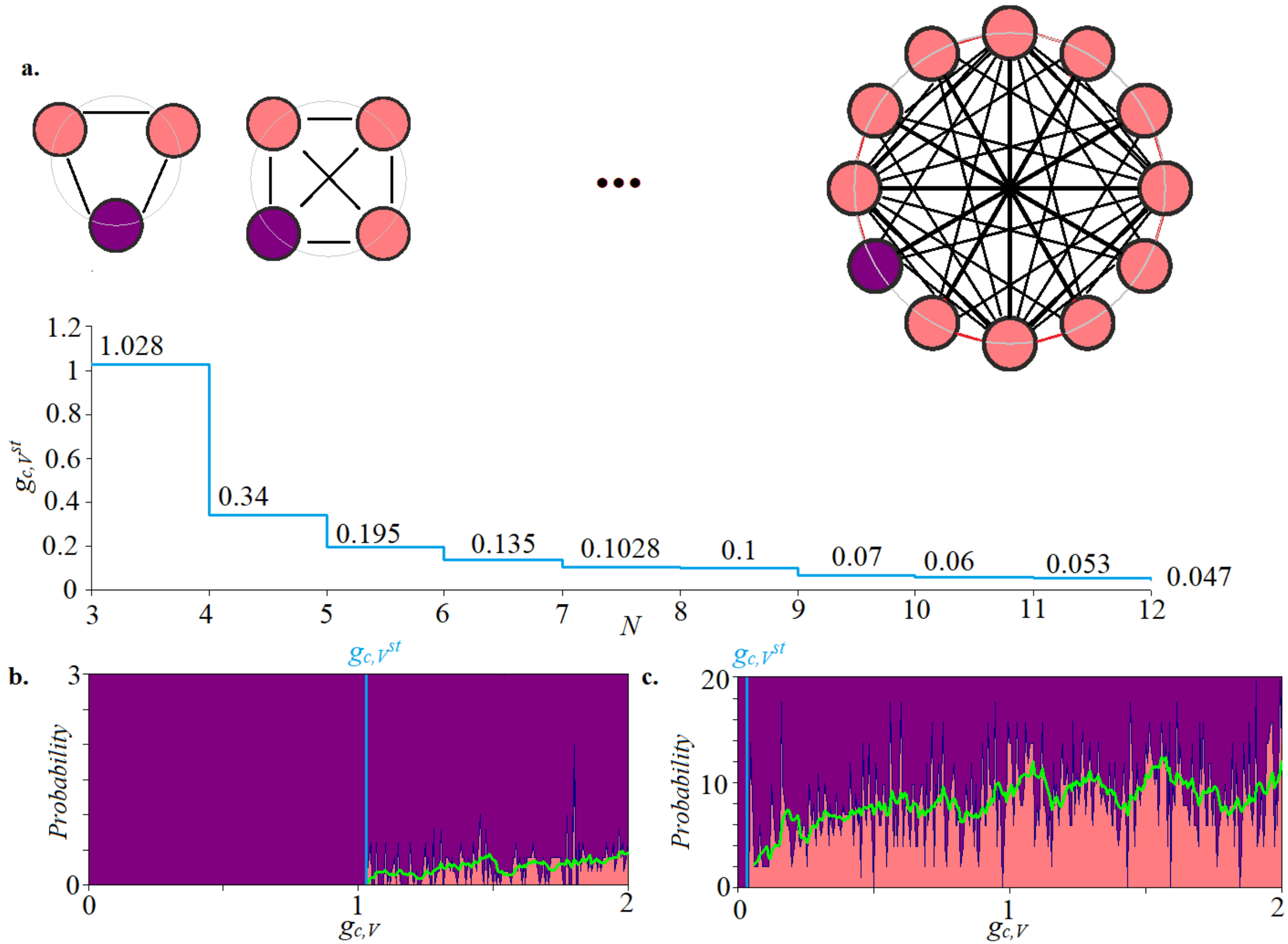}\\}
\caption{\textbf{Behavior of heterogeneous beta-cells population.}
 {\bf a.} (Top) Schematic representation of networks of increasing size, where only a single cell does not have a channel dis-regulations (red color of node responds to $k^{(i)}=1$, purple color to $k^{(i)}=0$) and (bottom) a corresponding plots of coupling strength thresholds for steady state stabilization in dependence on the size of the population ($N$).
 Probability of appearing of co-existing attractors in dependence on the coupling strength for the minimal population {\bf b.} of $N=3$ coupled cells, as well as for {\bf c.} $N=12$. Other parameters as in Fig.~\ref{fig:fig1}.
}
\label{fig:fig6}
\end{figure}

\begin{figure}
\centerline{
\includegraphics[scale=0.5]{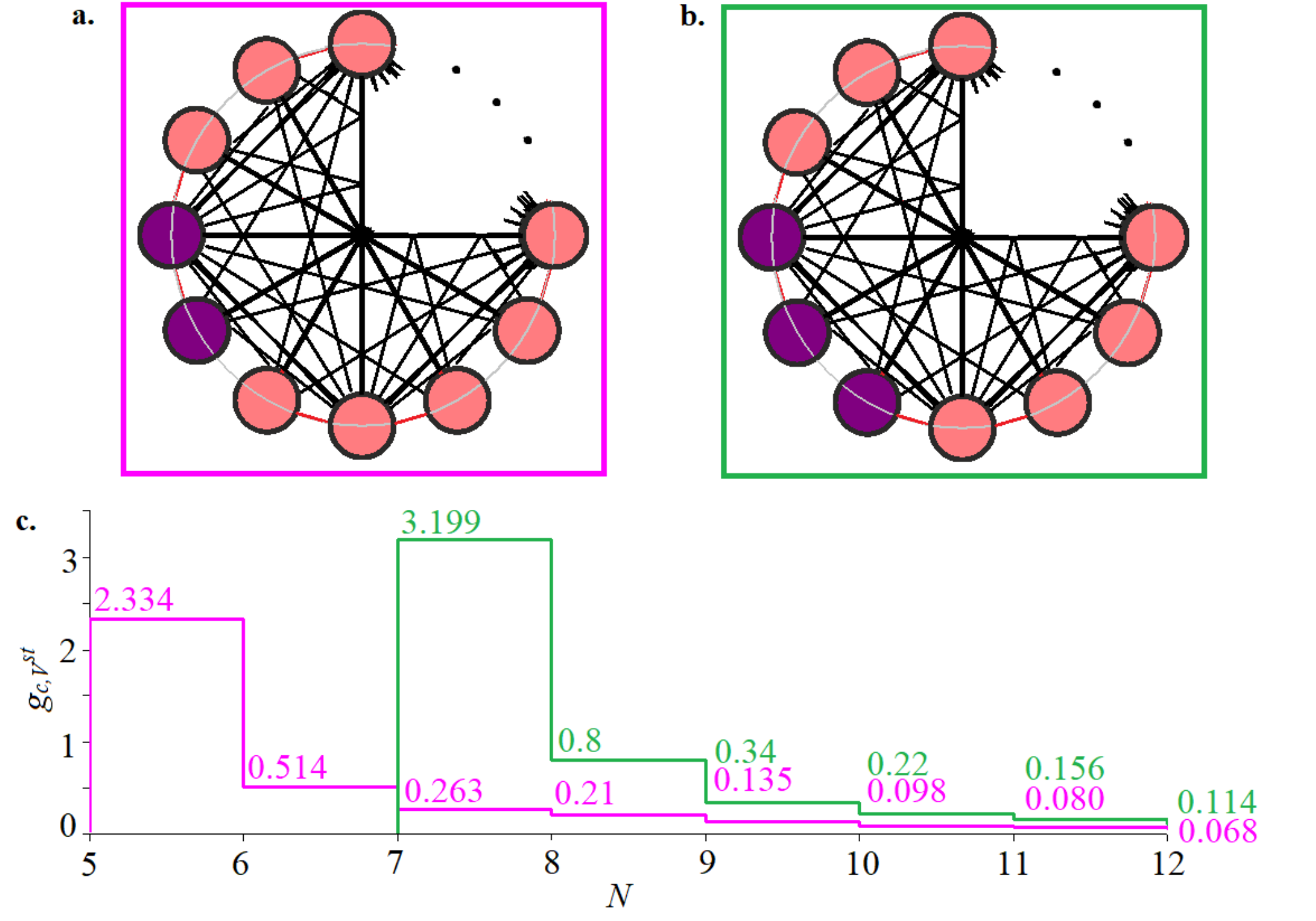}\\}
\caption{\textbf{Behavior of heterogeneous beta-cells population as depending on the number of cells without channel dis-regulations.}
 {\bf a,b.} Schematic representation  of networks of interacting cells (red color for $k^{(i)}=1$, purple color for $k^{(i)}=0$) and {\bf c.} corresponding plots depicting the thresholds of coupling strength for which the steady steady state is stabilized as a function of the size of the population ($N$). Other parameters as in Fig.~\ref{fig:fig1}.}
\label{fig:fig7}
\end{figure}

Bifurcation analysis showed that the mixed population does not display steady state dynamics over the whole $g_{c,V}$ interval (Fig.\ref{fig:fig4}b). This is contrary to the previous case (Fig.~\ref{fig:fig3}b) where silent state was dominant for intermediate $g_{c,V}$ values when bistability characterizes both cells, although the initial conditions of only one of them were poised in the stable node. A typical dynamics of the coupled mixed population model is mixed-mode oscillations, defined as complex oscillatory patterns where small-amplitude high-frequency oscillations alternate with large-amplitude long-period cycles \cite{a27}. Increasing the coupling strength between the two mixed oscillators induces transitions between different manifestations of the mixed-mode oscillations. For example, under low coupling strength, one oscillator displays small-amplitude whereas the other large-amplitude bursting dynamics (Fig.~\ref{fig:fig4}c). For intermediate coupling, different patterns of alternating between the small- and the large-amplitude bursting occur (Fig.~\ref{fig:fig4}d-f), to finally converge to almost synchronized bursting of both oscillators under strong coupling (Fig.~\ref{fig:fig4}g). The analysis therefore demonstrates that in a mixed population of two coupled cells, where only one of them has dis-regulation of a potassium channel, pathological silent dynamics is not stabilizied on the level of the population. This implies that deviations from the physiological bursting dynamics might be possible for a larger sub-population size that has channels dis-regulation.

\subsection{Effect of intercellular communication defects on the dynamics of heterogeneous beta-cells population}

To identify whether stabilizing pathological dynamics is a system-size effect, we investigated the dynamics of heterogeneous populations of increasing sizes, starting with a minimal population of $N=3$ cells, where $(N/2+1)$ of them can exhibit bistability between silent and bursting dynamics. The numerical simulations demonstrated that in this case there is a threshold of coupling strength for which the silent state can be stabilized on the population level (Fig.~\ref{fig:fig6}a).  However, the probability to observe this pathological, silent dynamics in the heterogeneous cell population is significantly smaller than the one in the homogenous population. Direct calculations from semi-random initial conditions as in Fig.~\ref{fig:fig3}b showed that for the minimal system of $N=3$ cells, this probability is smaller than $1\%$ (Fig.~\ref{fig:fig6}b). Subsequent increase of the population size by a cell that can display bistability  between a silent and bursting dynamics in turn lowered the coupling threshold for which the pathological, silent state was observed (Fig.~\ref{fig:fig6}a, bottom). The simulations also demonstrated that the probability to observe the pathological state increased with the increase of fraction of cells that have the channel defects. For example, for a population of $N=12$ cells, where $11$ of them have dis-functional channel increased to approximately $10\%$, almost over the full range of coupling strengths (Fig.~\ref{fig:fig6}c). Despite this increase in the probability to observe the pathological state, the population of heterogeneous cells can still cooperatively regulate the physiological bursting dynamics, even when the majority of the population has acquired dis-regulation in the channel's activity. This is contrary to a homogeneous population of cells with dis-regulation, where a coupling interval exists for which the probability to observe the pathological dynamics is $\sim100\%$ (Fig.~\ref{fig:fig3}b).

When the number of cells that lack dis-functional channels in the heterogeneous population is however increased, we find that the coupling threshold for which stabilization of the silent state is reached is subsequently increased. Indeed, in accordance with the previous finding that at least $(N/2+1)$ cells must have dis-functional channels for the silent state to be stabilized, the numerical simulations demonstrated that when $2$ or $3$ cells display only bursting dynamics, the minimal heterogeneous population size for which the silent state can be stabilized is $N=5$ (Fig.~\ref{fig:fig7}a,c) or $7$ (Fig.~\ref{fig:fig7}b,c) cells, respectively. The stabilization threshold scales with the number of cells displaying only bursting dynamics, being highest when the heterogeneous population has the largest number of cells exhibiting solely bursting dynamics. Similarly to Fig.~\ref{fig:fig6}a, a step-wise increase of the population by a cell that displays bistability between the silent and the bursting state also decreases the coupling threshold for stabilization (Fig.~\ref{fig:fig7}c).

\section{Discussion}

Intercellular communication dictates coordinated dynamics of cellular populations, and thereby determines cellular identity in terms of their function. For example, insulin secretion of pancreatic beta-cells is preceded by well-synchronized bursting activity of the population, where the secretion increases with the fraction of time the cells spend in the spiking state. In contrast, isolated cells do not burst and thereby do not secrete insulin effectively. Such dynamical behavior of single cells has been related to the decreased opening probability of the potassium channel, resulting in silent dynamics. We considered here the case when cells can exhibit silent dynamics, even when coupled in a beta-cells population. In particular, we addressed the question how channel defects on the level of single cells that alter the intercellular communication can affect the global dynamics of the population and under which conditions the dynamics will deviate from the physiological one. An altered dynamics would in turn affect the cellular function and thereby their identity.

The bifurcation analysis of a minimal homogeneous or heterogeneous populations of $N=2$ coupled cells, where respectively either both cells have channel defects and thereby can exhibit silent state or only one of them, demonstrated very different dynamical behavior on the population level. Whereas the homogeneous population exhibited dominance of the pathological silent state ($\sim100\%$) for a given coupling range despite the small basin of attraction on a single cell level, the minimal heterogeneous population was only characterized with bursting dynamics over the full coupling range. This suggested that the pathological state can only be stabilized as a system-size effect in a heterogeneous population of cells, when a significant proportion of cells in the population have channels with decreased probability of opening. The numerical simulations of larger heterogeneous populations on the other hand demonstrated that although the silent state could be indeed stabilized when the fraction of cells with channel dis-functions is increased, the bursting dynamics still remained a dominant dynamical behavior of the population, with prevalence of $\sim80-90\%$. Thus, the physiological population dynamics and the thereby associated cellular function, such as insulin production in the case of bursting dynamics of beta-cells can be reliably maintained, even in the presence of dis-functional channels on the level of single cells.

\section*{Acknowledgements}

N.S thanks the DAAD (Project Nr. 57440915) for the financial support of the scientific visit to MPI-MP in Dortmund, 2019.

\section*{Author contributions}

A.K. conceptualized the study. N.S. performed the numerical simulations and bifurcation analysis with help of A.K. N.S. and A.K. interpreted the results and wrote the manuscript.

\section*{Conflict of interest}

The authors declare that they have no conflict of interest.

\end{document}